\documentclass[journal=aamick,manuscript=article]{achemso}
\usepackage[version=3]{mhchem} 

\usepackage{graphicx,epstopdf}
\usepackage{amssymb,amsmath}
\usepackage{dcolumn}
\usepackage{bm}
\usepackage{color}
\usepackage{enumerate}
\usepackage{hyperref}
\usepackage{soul}

\usepackage{color}

\usepackage{comment}

\usepackage{cancel}

\usepackage[normalem]{ulem} 

\renewcommand{\vec}[1]{\mbox{\mathversion{bold}$#1$}}

\title[Brightening of the exciton...]
  {Strong light-matter coupling in carbon nanotubes as a route to exciton brightening}

\author{Vanik A. Shahnazaryan}
\affiliation{ITMO University, St. Petersburg 197101, Russia}

\keywords{carbon nanostructures, microcavity, exciton-polariton, dark exciton, photoluminescence
}

\author{Vasil A. Saroka}
\affiliation{Physics and Astronomy, University of Exeter, Stocker Road, Exeter EX4 4QL, United Kingdom}
\alsoaffiliation{Institute for Nuclear Problems, Belarusian State University, Bobruiskaya 11, 220030 Minsk, Belarus}

\author{Ivan A. Shelykh}
\affiliation{ITMO University, St. Petersburg 197101, Russia}
\alsoaffiliation{Science Institute, University of Iceland IS-107, Reykjavik, Iceland}

\author{William L. Barnes}
\affiliation{Physics and Astronomy, University of Exeter, Stocker Road, Exeter EX4 4QL, United Kingdom}

\author{Mikhail E. Portnoi}
\email{M.E.Portnoi@exeter.ac.uk}
\affiliation{ITMO University, St. Petersburg 197101, Russia}
\alsoaffiliation{Physics and Astronomy, University of Exeter, Stocker Road, Exeter EX4 4QL, United Kingdom}

\begin{document}

\begin{tocentry}
\centering
\includegraphics[width=0.65 \linewidth]{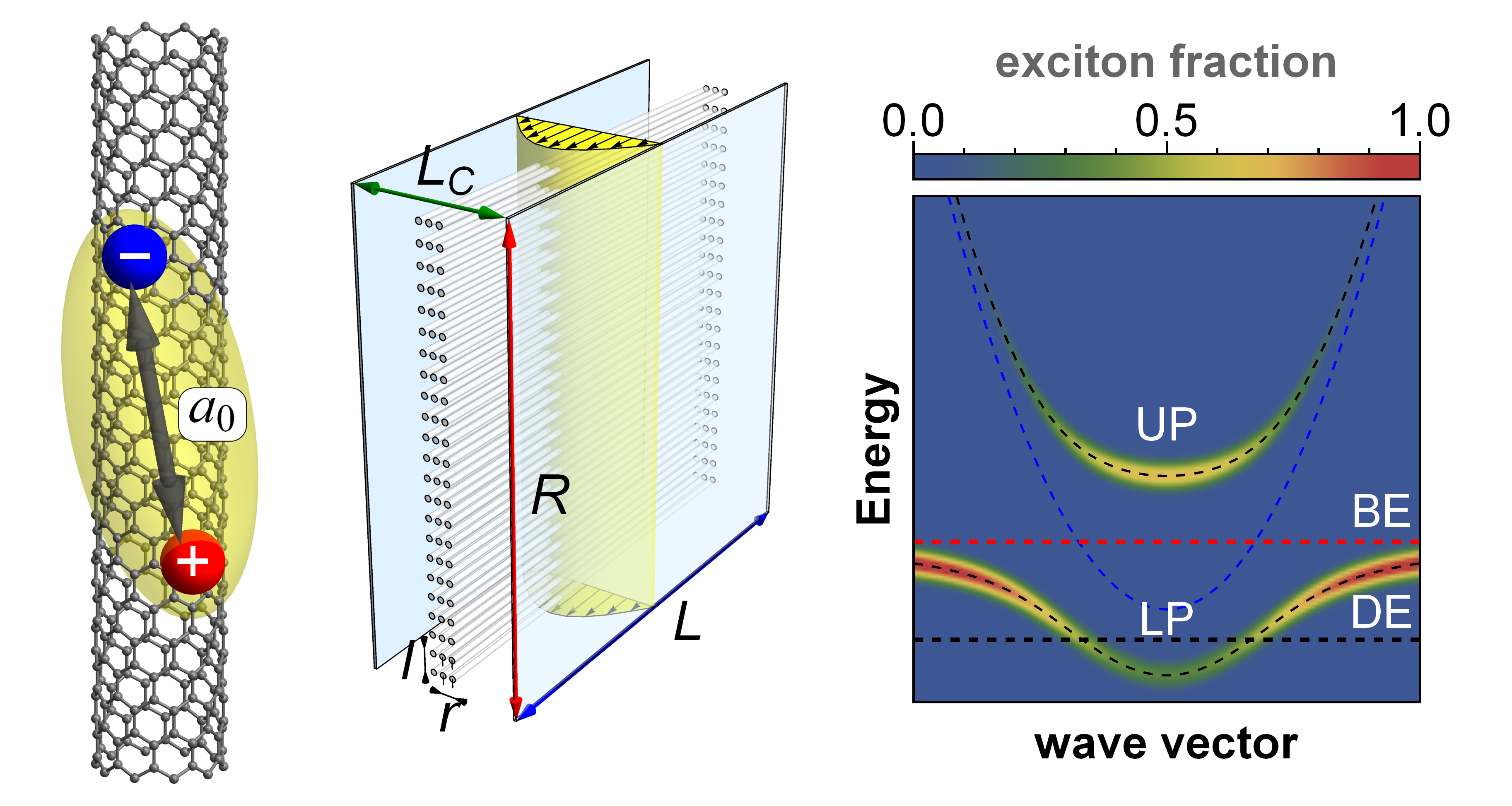}
\end{tocentry}



\begin{abstract}
We show that strong light-matter coupling can be used to overcome a long-standing problem that has prevented efficient optical emission from carbon nanotubes.
The luminescence from the nominally bright exciton state of carbon nanotubes is quenched due to the fast non-radiative scattering 
to the dark exciton state having a lower energy.
We present a theoretical analysis to show that by placing carbon nanotubes in an optical microcavity the bright excitonic state may be split into two hybrid exciton-polariton states, whilst the dark state remains unaltered.
For sufficiently strong coupling between the bright exciton and the cavity, we show that the energy of the lower polariton may be pushed below that of the dark exciton.
This overturning of the relative energies of the bright and dark excitons prevents the dark exciton from quenching the emission.
Our results pave the way for a new approach to band-engineering the properties of nanoscale optoelectronic devices.
\end{abstract}


\vspace{20pt}
When ensembles of emitters are placed inside an optical microcavity they may coherently exchange energy with each other via the cavity mode.
This exchange may lead to the formation of new hybrid light-matter states known as polaritons that are part-light and part-matter, this exchange process is known as strong coupling.
The hybrid states have energy levels that can be very different from those of the emitters from which they are formed, with remarkable consequences~\cite{Ebbesen_ACSaccounts_2016_49_2403}.
A slew of recent results, both theoretical and experimental, have shown that strong coupling can be used to modify chemical reaction rates~\cite{Hutchison_AngComm_2012_51_1592}, alter work-functions~\cite{Hutchison_AdvMat_2013_25_2481}, extend exciton transport~\cite{Orgiu_NatMat_2015_14_1123,Feist_PRL_2015_114_196402} and change molecular structure~\cite{Galego_PRX_2015_5_041022}.
Indeed, the effect of strong coupling in molecular systems is so pervasive and attractive that a new field known as polariton chemistry has recently emerged~\cite{Feist_ACSPhot_2017,Kowalewski_JPCL_2016_7_2050,Herrera_PRL_2016_116_238301}, itself part of a wider effort to use confined light to create and manipulate new states of matter~\cite{Mann_NatComm_2018_9_2194}. Strong coupling has been demonstrated across a wide spectral range~\cite{Shalabney_NatComm_2015_6_5981,Long_ACSPhot_2014_2_130}, for both solid and liquid media~\cite{George_JPCL_2015_6_1027}, and even encompasses biomolecules~\cite{Coles_NatComm_2014_5_5561,Dietrich_SciAdv_2016_2_e1600666}.
Here we show that these same ideas can be applied to provide a powerful technique with which to engineer the properties of emissive states by radically changing the properties of single-walled carbon nanotubes.

Single-walled carbon nanotubes (SWCNTs) were discovered more than two decades ago~\cite{Iijima1991} and possess remarkable electronic and optical properties~\cite{Charlier2007}. Of particular interest here is their potential as sources of light, SWCNTs are 1D excitonic systems possessing a strong electric dipole moment and thus have significant potential for optoelectronic technologies, and in quantum optics~\cite{He_NatMat_2018}.
Despite their potential, take-up of SWCNTs in optoelectronic applications has so far been limited~\cite{Park2013,DeVolder2013}.
This lack of take-up has been due to several important challenges. First, SWCNTs need to be produced with a specific chirality~\cite{SaitoBook1998} since the chirality (the way a nanotube is rolled from a graphene sheet) defines the SWCNT band structure and determines whether it exhibits metallic or semiconducting behavior. Further, the luminescence form SWCNTs can be severely quenched due to the presence of defects, such as nanotube ends, or nanotube bundles. The measurements on individual less defective tubes demonstrated enhanced photoluminescence efficiency (3-7$\%$ against $<1\%$ for defective samples)~\cite{Lefebvre2006,Carlson2007}. Moreover, the photoluminescence yield is sensitive to the local environment~\cite{Amori2018}. Thus, the issues related to fabrication imperfections hinder the utilization of optical properties of SWCNTs~\cite{He_NatMat_2018}. However, recently impressive progress has been achieved in the synthesis and post-processing of carbon nanotubes~\cite{Janas2018}. The chirality selective growth of metallic (6,6) SWCNTs has been demonstrated~\cite{Sanchez-Valencia2014}. Multi-column gel chromatography has been shown to be able to separate up 12 different single-chirality enantiomers with up to 97$\%$ of purity~\cite{Wei2016}. Furthermore, the single chirality tubes have been assembled into well-aligned films via controlled vacuum filtration~\cite{Titova2015,He2016,GaoArxiv2018}. These achievements give confidence that the problems of SWCNT sample fabrication will be gradually solved with improvements in experimental techniques.

The third challenge that must be overcome to realize the potential of SWCNTs in optoelectronics is photophysical in nature; the luminescence associated with nanotube excitons is dramatically suppressed, even at room temperature~\cite{Shaver2007}. The reduction of luminescence efficiency can be explained by the presence of dark exciton states. Along with a high energy so-called K-momentum dark exciton state~\cite{Blackburn2012,Blackburn2016,Amori2018}, there is a non-radiative dark exciton state, having a significantly lower energy than the radiative bright excitonic states~\cite{Spataru2004,Perebeinos2004,Perebeinos2005,Scholes2007}. Bright excitons relax towards this dark state with the consequence that non-radiative decay dominates over the desired radiative relaxation channel ~\cite{Avouris2007,Avouris2008}. In the present work we focus on overcoming the photophysical limitation imposed by the presence of the low-lying dark state.

A number of methods to enhance the luminescence efficiency have already been investigated.
One approach is chemical functionalization~\cite{Kilina2012}, but it is questionable whether it can be achieved with the current state of technology.
A second approach seeks to make use of the Purcell effect to increase the radiative rate of the bright exciton by placing nanotubes in a very low volume microcavity\cite{Miura_NatComm_2014_5_5580,Luo_NatComm_2017_8_1413,Jeanet_NL_2017_17_4184}. Although interesting, especially from a single photon source perspective, the use of very low cavity volumes precludes this approach being adopted in optoelectronic applications.
A third approach exploits high-power pulsed-laser irradiation at room temperature~\cite{Harutyunyan2009}.
However, irradiation brightening requires strong laser fields and these lead to irreversible damage to the tubes. 

Perhaps the most impressive approach adopted to date is that of magnetic brightening~\cite{Shaver2007,Shaver2007a,Srivastava2008}; it has already been used to demonstrate that bright states exist.
An external magnetic field breaks time-reversal symmetry and this changes the nanotube band structure. 
In these experiments the degeneracy between the two minima in the SWCNT energy spectrum is lifted by applying a magnetic field, the applied field leads to a shrinkage of the direct band gap at one of the {\bf K}-points (Dirac points).
This change to the band structure moves the bright exciton state below the dark exciton state. The integrated luminescence at low temperatures is found to increase five-fold~\cite{Shaver2007}, without any significant change in the transition matrix elements or changes to the bright exciton structure being required.

However, due to the small diameter of SWCNTs, achieving the required flux density within a tube for magnetic brightening requires magnetic fields of about $5$-$50$ T~\cite{Shaver2007a,Srivastava2008} or pulsed fields of up to 190 T~\cite{Zhou2014}. 
Nevertheless, despite lacking practical utility, magnetic brightening is important because it shows that if we are able to shift energetically the bright state below the dark one, the luminescence from the tube increases many fold; the physical mechanism by which the energy of the lowest bright state is reduced is not important.  


In this paper we propose an elegant practical alternative to overcome the problem of non-radiative relaxation of what would otherwise be a bright excitonic state of SWCNTs.
Our approach involves depressing the energy of the `bright' state so that it is below the energy of the dark state that causes the non-radiative leakage from bight state, thereby turning the luminescence quenching process off.
We do this by placing an ensemble of SWCNTs within the confined electromagnetic mode of an optical microcavity. Strong light-matter coupling between the cavity mode and the `bright' exciton state leads to the formation of two exciton-polariton branches, the upper polariton (UP) and the lower polariton (LP).
For a suitable design the energy of the lower polariton is pushed down below that of the dark exciton.
The lowest-energy dark exciton, which is formed by an electron and a hole belonging to different valleys, has zero oscillator strength and thus does not interact with the cavity mode.

\begin{figure}
  \centering
  \includegraphics[width=0.78 \linewidth]{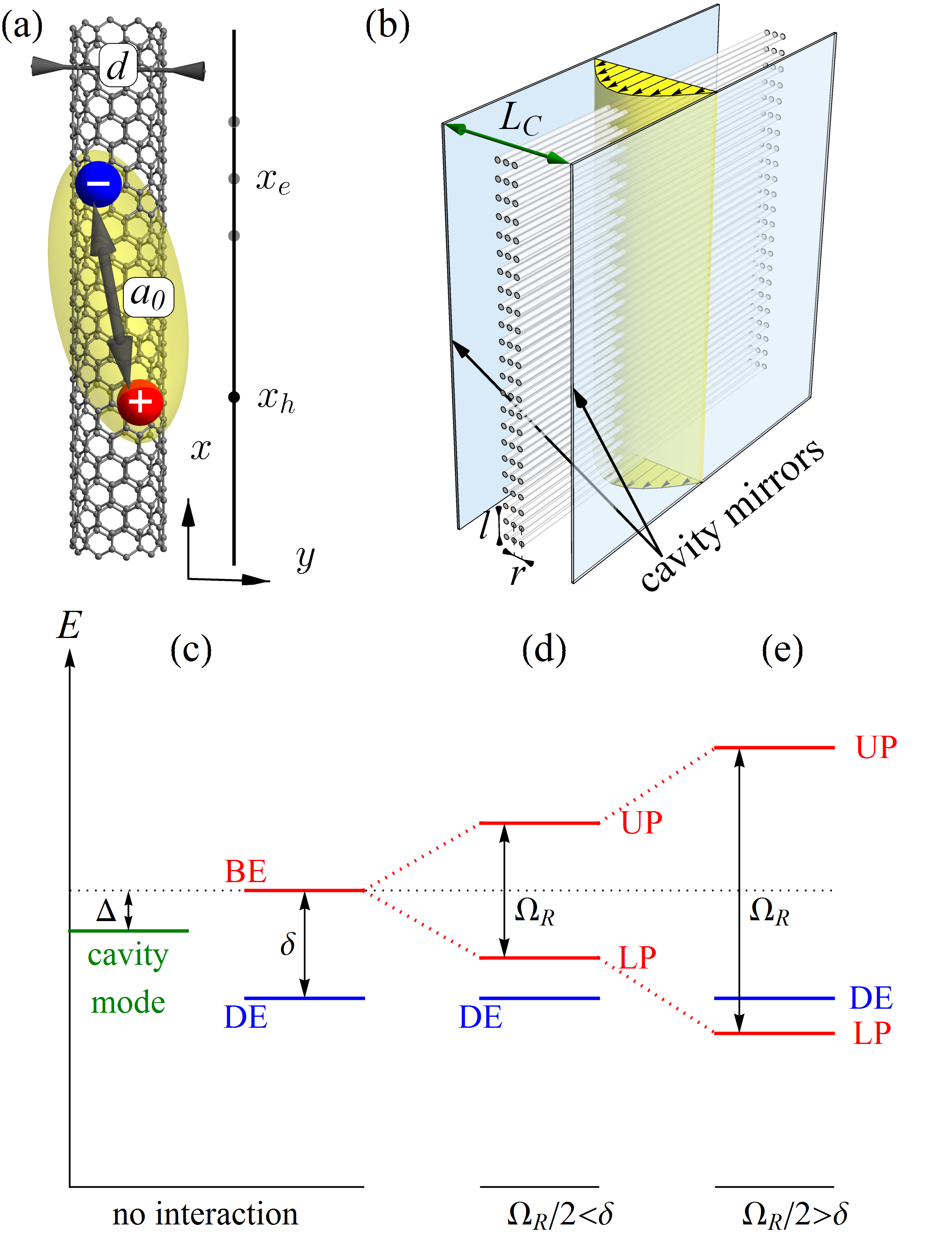}
  \caption{(a) A sketch of a semiconductor carbon nanotube, of diameter $d$, containing an exciton (bound electron-hole pair) and its 1D mathematical representation.
The characteristic size of the electron-hole pair is the excitonic Bohr radius $a_0$.
(b) The geometry of the structure considered in the present paper: a planar microcavity with an array of aligned carbon nanotubes embedded in a position such that the electric field of the confined cavity mode (shown in yellow) reaches its maximum at the centre of the cavity. Here $l$ and $r$ denote the in-plane distance between nanotubes and distance between the layers of nanotubes, respectively.
Strong coupling between bright excitons and confined photons leads to the formation of hybrid polariton modes in the system (c-e).
(c) The energy levels of the uncoupled bright and dark excitons, and the cavity mode (green line).
(d) Strong light-matter coupling leads to the hybridization of the bright exciton (BE) and cavity (photonic) mode, resulting in the appearance of upper (UP) and lower (LP) polaritons separated by the value the vacuum Rabi splitting $\Omega_R$. Here $\Omega_R/2<\delta$, where $\delta$ is splitting between bright and dark excitons (DE), so that the lowest exciton state, i.e. the ground state, is the dark state.
(e) Here $\Omega_R/2>\delta$ so that the energy of the lower polariton sinks below the energy of the dark exciton; the lowest exciton state, i.e. the ground state, is now a bright state, the LP.}
\label{sketch}
\end{figure}

The key features of our concept are sketched in Fig.~\ref{sketch}. In Fig.~\ref{sketch}(a) we show an exciton associated with a carbon nanotube. 
In Fig.~\ref{sketch}(b) we show an ensemble of nanotubes located in the central plane of an optical microcavity, the electric field distribution of the cavity mode is also indicated.
In Fig.~\ref{sketch}(c-e) we show the effect of strong coupling on the exciton energy levels.
In Fig.~\ref{sketch}(c) the energy levels of the uncoupled excitons and of the cavity are shown.
In the absence of a microcavity the bright exciton state has a higher energy than the dark exciton state, they are separated by a frequency we label $\delta$.
In Fig.~\ref{sketch}(d) the effect of strong coupling is indicated, the bright exciton is split into an upper and lower polariton branch, UP and LP respectively, but the extent of this vacuum Rabi splitting, $\Omega_R$ is less than the intrinsic bright-dark exciton splitting of $\delta$.
In Fig.~\ref{sketch}(e) the strong coupling is much stronger so that now $\Omega_R/2 > \delta$ and the energy level ordering of the no-cavity bright and dark states has been overturned.

The strong-coupling regime between light and matter is achieved when the characteristic coupling rate (energy) of the interaction between material excitations and the electromagnetic cavity mode exceeds the damping rates (energies) associated with losses in the system~\cite{Microcavities}.
Recent work suggests that our proposal may be achieved in experiment; strong coupling of ensembles of carbon nanotubes inside a planar optical microcavity has been the subject of several recent reports~\cite{Graf2016,Zakharko2016,Graf2017,Gao2018}. Indeed, carbon nanotubes share many attractive features for strong coupling that are associated with other organic quantum emitters, notably the intrinsically large binding energies and oscillator strengths of the excitons involved~\cite{KCohen2010,KCohen2013,KCohen2014,Plumhof2014}.

The characteristic splitting between dark and bright excitons is usually several tens of meV \cite{Spataru2005}, thus brightening of the ground state requires the vacuum Rabi splitting, $\Omega_R$, to be similar in extent.
Below we demonstrate that this criterion can be met in experimentally realizable configurations, similar to those presented in recent experiments~\cite{Graf2016,Zakharko2016,Graf2017}.\\

Our task is to evaluate the energy of the polariton states and to explore the conditions that will need to be achieved for the lower polariton to have an energy that is lower than that of the dark exciton. In what follows we outline our approach; we begin by discussing the dispersion of the polariton modes.\\

\section*{Polariton Dispersion}
The wavevector dependent energies of the polaritons are found from the eigenvalues of the polariton Hamiltonian~\cite{Microcavities},
\begin{equation}
\label{polar_Ham}
		\hat{H}=\begin{pmatrix}
		E_C(k) & g(k)/2 \\
        g(k)/2 & E_X    \\
	\end{pmatrix},
\end{equation}
where, for simplicity, we have neglected the exciton dispersion because the effective mass of an exciton, $m_X$, is much greater than that of the photon effective mass in the cavity, $m_C$, i.e. $m_X \gg m_C$. The eigenvalues are found to be,
\begin{equation}
\label{eq:polariton_energies}
E_{\mathrm{UP(LP)}}=\frac{E_X+E_C(k) \pm \Omega_R(k)}{2},
\end{equation}
where the upper polariton, $E_{UP}$, is given by taking the $+$ sign and the lower polariton, $E_{LP}$, by taking the $-$ sign.
$E_X$ is the unperturbed exciton energy (derived below), whilst $E_C$ is the unperturbed cavity mode energy whose energy in the parabolic approximation is given by,
\begin{equation}
\label{eq:cavity_mode}
E_C(k) = E_C(0)+{\hbar^2 k^2}/{(2 m_C)},
\end{equation}
with: $m_C=2\pi \hbar \sqrt{\varepsilon}/(c\lambda_C)$ standing for the photon effective mass;
$\lambda_C=2L_C$ is the cavity resonance wavelength, with $L_C$ the cavity length; $k$ is the in-plane wavevector; $c$ the speed of light; and $\varepsilon$ the relative permittivity of the medium inside the cavity.
Finally for equation \ref{eq:polariton_energies}, the Rabi splitting, $\Omega_R(k)$, is defined as,
\begin{equation}
\label{eq:Rabi_splitting}
\Omega_R(k)=\sqrt{(E_X-E_C(k))^2+g^2(k)},
\end{equation}
where $g(k)$ is the cavity-exciton interaction coupling rate for the nanotube ensemble.

To calculate the desired wavevector dependent energies of the polariton modes, Eq. (\ref{eq:polariton_energies}), we thus need to know the energy of the unperturbed cavity mode $E_C$, the unperturbed exciton energy $E_X$, and the cavity-exciton interaction coupling rate $g(k)$.
The cavity energy $E_C$ is straightforward and has already been given in equation \ref{eq:cavity_mode}. Since the two remaining quantities, $E_X$ and $g(k)$ both depend on the exciton wavefunction we look at the exciton wavefunction next.\\

\section*{Exciton wavefunctions}
The allowed wavefunctions are obtained by seeking solutions of the Schr\"{o}dinger equation subject to a suitable potential.
Previously a number of different theoretical approaches to treat excitonic states in carbon nanotubes have been adopted, with varying degrees of complexity, these include: numerical results obtained in the framework of $\vec{k}\cdot \vec{p}$ effective mass models~\cite{Ando1997,Kane2003,Kane2004,Ando2006c,Uryu2007,Ando2009}, first principles calculations~\cite{Chang2004,Spataru2004,Spataru2005}; solutions of the Bethe-Salpeter equation in the tight- binding approximation~\cite{Perebeinos2004,Perebeinos2005,Jiang2007}; Pariser-Parr-Pople models~\cite{Zhao2004}; and variational approaches~\cite{Pedersen2003,Maultzsch2005}.
In addition, analytical solutions of two-particle problems have been found for some classes of model interaction potentials~\cite{Hartmann2011,Downing2014a}. 
In the present work we use a combination of the effective-mass and envelope-function approaches along with the low-energy zone-folding tight-binding model.
This choice allows us to obtain semi-analytical results for excitonic energies and wavefunctions which can be easily utilised in a well-established formalism for treating exciton-polaritons.
In this case the electron-hole interaction in the tube can be approximated by a shifted Coulomb potential~\cite{Loudon1959,Wang2005},
\begin{equation}
\label{eq:CoulombPotential}
V(x)=-\frac{1}{4\pi\varepsilon_0}\frac{e^2}{\varepsilon_C (|x|+ \gamma d)},
\end{equation}
where: $x = x_e - x_h$ is the distance between electron, $x_e$, and hole, $x_h$, positions on the nanotube axis; $d$ is the diameter of the nanotube, as shown in Fig.~\ref{sketch}(a); $e$ is the elementary charge; $\varepsilon_0$ is the permittivity of free space; $\varepsilon_C =2$ is the static dielectric constant of the carbon nanotube~\cite{Zhao2004,Jiang2007}; and $\gamma$ is a cut-off scaling parameter.
The Coulomb interaction~\eqref{eq:CoulombPotential} in general mixes all of the sub-bands, but for narrow tubes (with diameters $d<2$~nm) the sub-band extrema are well separated in energy and a two-band model can be used. When $ d \ll a_0$, where $a_0$ is the characteristic size of an exciton, the problem becomes equivalent to the phenomenological analytical model of a 1D exciton developed by Loudon~\cite{Loudon1959} and successfully applied to the description of excitons in semiconducting quantum wires~\cite{Banyai1987,Ogawa1991,Ogawa1991a} and in semiconducting carbon nanotubes~\cite{Wang2005}.

To determine the cut-off scaling parameter $\gamma$ in Eq.~\eqref{eq:CoulombPotential} we analysed numerical and experimental data on exciton binding energies, $E_b$, from Refs.~\cite{Zhao2004,Chang2004,Maultzsch2005,Pedersen2003}.
In Loudon's model, the binding energy of the exciton is related to the exciton quantum number $\alpha$, which can be fractional due to a quantum defect, via~\cite{Loudon1959},
\begin{equation}
\label{eq:ExcitonBindingEnergy}
    E_b =  -\dfrac{\hbar^2}{2 \mu a_0^2 \alpha^2} \, .
\end{equation}
where $a_0 =  4 \pi \varepsilon_0 \varepsilon \hbar^2/(\mu e^2)$ is the excitonic Bohr radius and $\mu=m^{\ast}/2$ is the reduced mass of an electron-hole pair.
In the approximation used here the effective masses of an electron and a hole are equal, and within the tight-binding model can be estimated as  $m^{\ast}=(2 \hbar)/(3 d v_F)$, with $v_F$ being the Fermi velocity in graphene.
Combining the data on exciton binding energies from the various sources noted above, i.e. ~\cite{Zhao2004,Chang2004,Maultzsch2005,Pedersen2003}, we can extract $\alpha$'s from Eq.~\eqref{eq:ExcitonBindingEnergy} and average them to yield a value of $\alpha_{\text{av.}} = 0.64(9)$.
With this value for the exciton quantum number we obtain (via Eq.~(3.22) in Ref.~\cite{Loudon1959}) a value for the cut-off scaling parameter of $\gamma = 0.87(5)$. 
We note that this value is different from that in Ref.~\cite{Wang2005}, where $\gamma$ was fixed at $0.3$ and the static dielectric constant was used as an adjustable parameter in Loudon's model.
Here we employ a slightly different approach to the problem since we assume that the static dielectric constant in all experiments should be dominated by the carbon nanotube dielectric constant $\varepsilon_C$.

We use our value of $\alpha_{\text{av.}}$ and the resulting value of $\gamma$ to calculate the excitonic wave function, which is expressed in terms of the Whittaker function of the second kind, $W_{\alpha,1/2}$, as~\cite{Loudon1959},
\begin{equation}
\label{eq:ExcitonWavefunction}
    \psi_{\alpha}(x) = N_{\alpha} W_{\alpha,1/2}(2(|x|+ \gamma d)/\alpha a_0),
\end{equation}
with $N_{\alpha}$ being a normalization constant.

Fortunately the ratio $d/a_0$ is universal for all diameters of tube, this universality can be seen by calculating the excitonic Bohr radius, $a_0$, which is given by $a_0 =  4 \pi \varepsilon_0 \varepsilon_C \hbar^2/(\mu e^2)$ where $\mu=m^{\ast}/2$ is the reduced mass of an electron-hole pair.
The effective masses of an electron and a hole are equal in our approach, and within the tight-binding model can be estimated as $m^{\ast}=(2 \hbar)/(3 d v_F)$, with $v_F$ being the Fermi velocity in graphene and $d$ the nanotube diameter.
Making use of the relative permittivity for the carbon nanotubes of $\varepsilon_C = 2$ and taking the Fermi velocity to be $10^6~ms^{-1}$ this results in $d/a_0 = 0.38$.
Now that the wavefunctions can be calculated the next task is to determine the exciton energies.\\

\section*{Exciton energy $E_X$}
The energy corresponding to photo-excitation of an exciton is defined as $E_X=E_g-E_b$ where $E_g$ is the energy of the bandgap and $E_b$ is the exciton binding energy. The exciton binding energy can be calculated using the model outlined above, whilst the SWCNT bandgap energy is easily calculated using a standard approach given in Supporting Information part A. All that remains to determine the polariton dispersion is to evaluate the cavity-exciton interaction coupling rate.\\

\section*{Cavity-exciton interaction coupling rate $g(k)$} 

The interaction rate between cavity eigenmode and the exciton mode of single SWCNT is \cite{Burstein}
\begin{equation}
\label{g_0}
g_{0}(k)=\sqrt{\frac{E_C(k)}{\varepsilon \varepsilon_0 V}} \sqrt{L} |\psi_{\alpha} (x=0)| d_{cv},
\end{equation}
where $d_{cv}$ is the interband transition dipole matrix element, $L$ is the length of SWCNT, and $V=LL_C Nl$ is the cavity volume. Here $L_C$ is the distance between mirrors, $l$ is the inter-tube distance, and $N$ is the number of SWCNT within one layer. We next note that in long wavelength limit $L_C\gg n r$ (where where $n$ is the number of layers of nanotubes in the cavity and $r$ is the distance between layers) all the exciton modes interact equally  with cavity mode, meaning that overall light-matter coupling rate of the structure reads~\cite{Fox_QUANT}
\begin{equation}
g(k)=\sqrt{nN}g_0(k),
\end{equation}
 Inserting the value of $g_0$ given by Eq.~(\ref{g_0}), we end up with the result

%
\begin{equation}
	\label{Rabi-2}
  	g(k)=\sqrt{\frac{n}{l}}\sqrt{\frac{E_C (k)}{\varepsilon \varepsilon_0 L_C}}  d_{cv} \left|\psi_{\alpha}(x=0)\right|.
\end{equation}
%

For the permittivity $\varepsilon$, the cavity is assumed to be filled by the polymer material PFO-BPy with effective permittivity $\varepsilon=2.8$~\cite{Graf2016}, although for the characterization of the excitonic states we use the pure carbon dielectric constant $\varepsilon_{C}=2$.
Finally, the quantity $d_{cv}= 3ed/4$ is the interband transition dipole matrix element and is derived in Supporting Information part A.\\

\section*{Results and Discussion}
Our calculated bandgap energy $E_g$, exciton binding energy $E_b$, and zero in-plane wavevector coupling strength $g(0)$ for excitons in SWCNTs of different chiralities are presented in Table~\ref{tbl}. These parameters form the basis for the results reported below.

\begin{table}
\centering
    \begin{tabular}{c c c c c c}
        \hline
        SWCNT & d (nm) & $\mu$ ($m_0$) & $E_g$ (eV) & $E_b$ (eV) & $g(0)$ (meV)   \\ \hline
        (7,0) & 0.548  & 0.072         & 1.57       & 0.564       & 68.69         \\ \hline
        (6,2) & 0.565  & 0.069         & 1.53       & 0.555       & 68.43         \\ \hline
        (8,0) & 0.626  & 0.062         & 1.38       & 0.499       & 64.81         \\ \hline
        (6,4) & 0.683  & 0.057         & 1.26       & 0.459       & 60.99         \\ \hline
        (9,1) & 0.747  & 0.052         & 1.15       & 0.419       & 58.66         \\ \hline
        (6,5) & 0.747  & 0.052         & 1.15       & 0.419       & 58.66         \\ \hline
        (8,3) & 0.771  & 0.051         & 1.12       & 0.411       & 57.86         \\ \hline
       (10,0) & 0.783  & 0.050         & 1.10       & 0.403       & 57.05         \\ \hline
        (7,5) & 0.818  & 0.048         & 1.05       & 0.386       & 55.91         \\ \hline
       (11,0) & 0.861  & 0.045         & 1.00       & 0.362       & 54.56         \\ \hline
        (7,6) & 0.882  & 0.045         & 0.98       & 0.362       & 54.08         \\ \hline
        (9,4) & 0.903  & 0.043         & 0.95       & 0.346       & 52.85         \\ \hline
       (13,0) & 1.020  & 0.038         & 0.85       & 0.306       & 50.74         \\ \hline
       (14,0) & 1.100  & 0.036         & 0.79       & 0.290       & 48.96         \\ \hline
    \end{tabular}
    \caption{Nanotube diameters $d$, exciton effective masses $\mu$, bandgaps $E_g$, exciton binding energies $E_b$ and coupling constants between the excitons and confined cavity mode $g$ for SWCNTs of different chiralities.
The values of $g$ were calculated for the case of the resonance between photons and bright excitons for an inter-tube separation of $3$~nm.
We note that within a low-energy tight binding model and neglecting the trigonal warping effect~\cite{Saito2000}, the excitonic properties are defined solely by the diameter of the nanotube.
Thus the identical parameters of nanotubes of (9,1) and (6,5) chirality are explained by the fact, that they have the same diameter, defined by $d=(a/\pi)\sqrt{n^2+m^2+nm}$, where $a=2.46$~\AA{ }is the graphene lattice constant, and $n$, $m$ denote chiral indices. }
    \label{tbl}
\end{table}

We next discuss aspects of the cavity design. To reach the strong coupling regime requires the light-matter interaction to be strong enough to overcome all dissipative processes in the system. The emission linewidth of the SWCNT emission is $\sim 20$ meV, whilst cavities with a Q of $>$ 100 are now routine ~\cite{Microcavities} so that a cavity linewidth of $<$ 20 meV is not a problem; strong light-matter coupling thus requires $g(k)\geq 20$ meV, something that has already been achieved \cite{Graf2016}. 
For the sake of simplicity, here we consider the case when the first cavity eigenmode is close to resonance with the excitonic transition.
For this mode the amplitude of the electric field is maximal at the centre of the cavity and only the tubes located close to the cavity centre will efficiently participate in light-matter coupling.
We thus consider a structure with a few layers of densely packed parallel SWCNTs (see Fig. \ref{sketch} (b)).
It is possible to fabricate high density arrays of carbon nanotubes using a spontaneous alignment technique~\cite{He2016}, with a minimum distance between the nanotubes down to 1 nm.
However, to avoid the effects of  band structure modification due to electron hopping between adjacent tubes, in the current work we limit ourselves to the case of less dense ensembles where $l>3$ nm, for which inter-tube electron hopping can be neglected.
As we demonstrate below, ground state exciton brightening can be achieved even in this diluted regime.

Since the resonant wavelength of the cavity mode satisfies the condition $\lambda_C \gg n r$, where we recall that $r$ is the distance between the layers of nanotubes and $n$  is the number of layers, one can safely neglect the variation in cavity mode field strength across the nanotubes for a few layer system. To be specific, hereafter we choose $n=10$.

\begin{figure*}
  \centering
  \includegraphics[width=1.0 \linewidth]{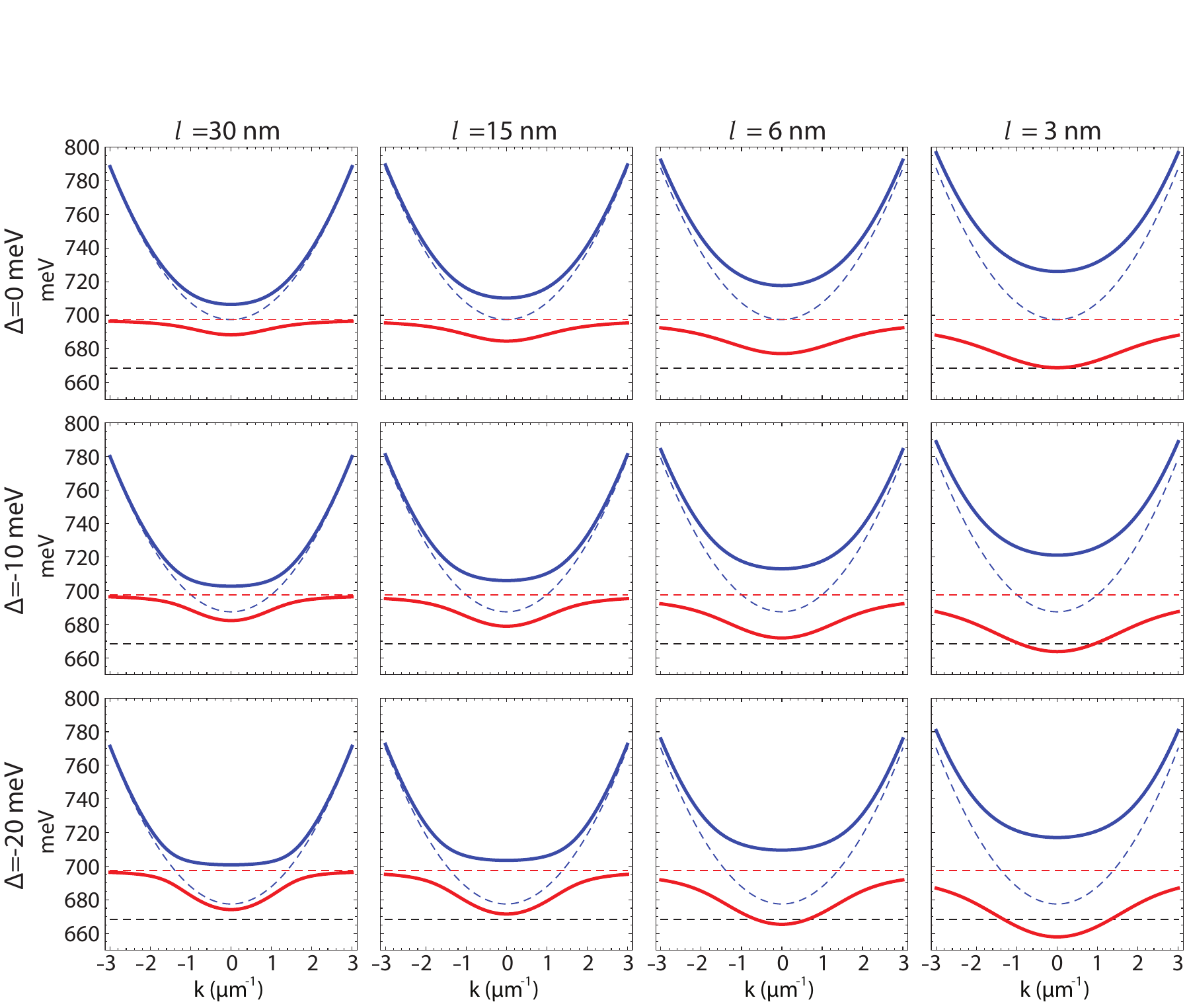}
  \caption{The dispersion of the polariton modes versus those of bare excitons and cavity photons for different detunings (from top to bottom) and for different nanotube separation distances (from left to right) for the case of a (10,0) SWCNT.
The dashed lines correspond to the bare photon (blue), bright exciton (red) and dark exciton (black) modes, solid red and blue lines correspond to the upper and lower polaritons, respectively. A decrease of the inter-tube separation is seen to result in an increase of the Rabi splitting. The lower polariton mode moves down in energy as the inter-tube separation is decreased and, below a certain separation, crosses the dark exciton. In this regime the brightening of the ground state of the system takes place.}
  \label{dispersion}
\end{figure*}

Using the information and parameters given above we are now in a position to calculate the properties of the polariton modes.
The dispersion of these modes are shown in Fig.~\ref{dispersion} for the case of (10,0) nanotubes and for different levels of cavity-exciton detuning $\Delta=E_C(0)-E_X$.
(In practice the detuning may be set by adjusting the geometry of the cavity, in particular the cavity length $L_C$.)
For the chosen chirality the energy splitting between bright and dark excitons $\delta=E_X-E_X^{\mathrm{dark}}=29$ meV \cite{Spataru2005}.
In Fig.~\ref{dispersion} we see that an increase in tube concentration leads to an enhancement of the light-matter coupling, pulling down the energy of the lower polariton branch. For the maximum tube density considered, corresponding to a separation distance of $l_{min}=3$ nm, the coupling strength is about $g \sim  55 $~meV.
We note that both the square root dependence of the interaction strength on the SWCNT density (see. Eq.~(\ref{Rabi-2})) and coupling values are in good agreement with existing experimental results \cite{Graf2016}.

\begin{figure}
  \centering
  \includegraphics[width=1.0 \linewidth]{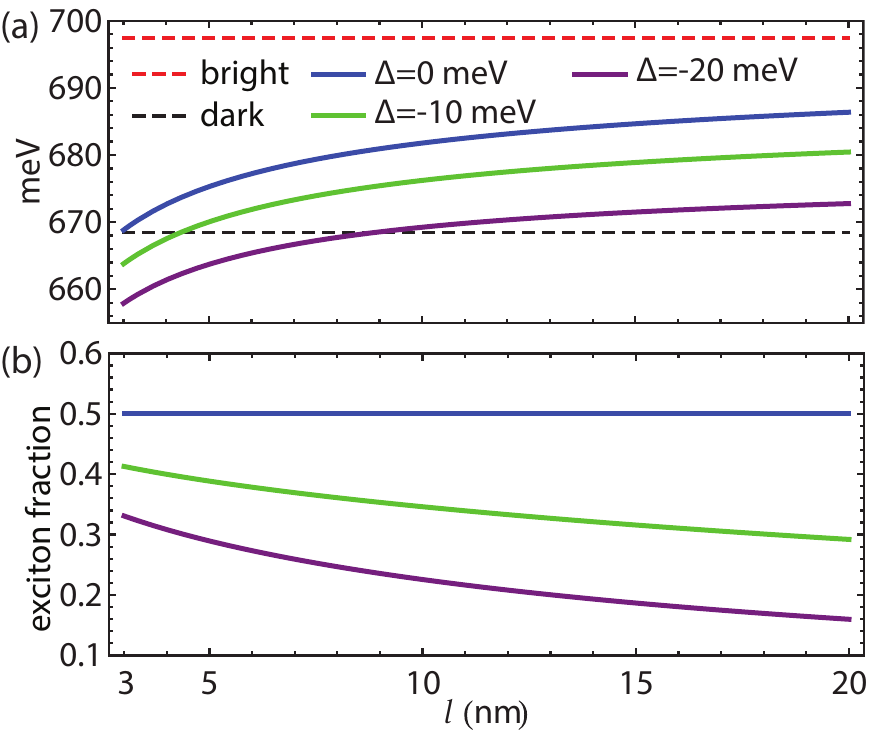}
  \caption{(a) The position of the bottom of the lower polariton branch as function of the inter-tube distance $l$, for the case of a (10,0) SWCNT, for different detunings versus dark exciton energy (dashed black line). For small values of $l$ the lower polariton lies below the dark exciton and the ground state brightens. (b) Exciton fraction in the lower polariton mode as a function of the inter-tube distance $l$ for the case of a (10,0) SWCNT for different detunings.}
  \label{bright}
\end{figure}

As one can see from Fig.~\ref{dispersion}, for distances between the nanotubes close to $l_{min}=3$ nm, the energy of the lower polariton can reach the value of the energy of the dark exciton state.
Introducing a non-zero detuning between excitonic and photonic modes, as shown in the lower panels of the figure, can pull the lower polariton state down further so that it is below the dark exciton state.
Fig.\ref{bright} shows the position of the lower polariton energy and the excitonic fraction of the lower polariton as a function of the distance between the nanotubes for various values of detuning.
Importantly, the exciton fraction of the lower polariton remains significant, even in the presence of substantial detuning; this is a direct consequence of the large values of the Rabi splitting $\Omega_R$.
This is important because, even though the lower polariton has been down-shifted in energy enough to take it below the dark exciton, the lower polariton still carries a significant excitonic character and will thus still be a good emitter of light.

\begin{figure}
  \centering
  \includegraphics[width=1.0 \linewidth]{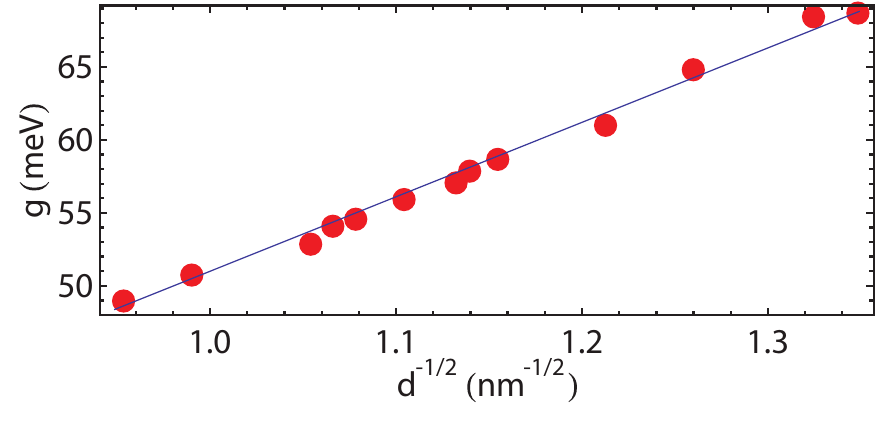}
  \caption{The dependence of the light-matter coupling constant on the nanotube diameter. The decrease of diameter leads to the increase of the light matter coupling. Red points correspond to our numerical data, the black curve is a numerical fit corresponding to a $d^{-1/2}$ scaling. The tube separation distance was chosen as $l=3$~nm and the detuning $\Delta=0$.}
  \label{g_d}
\end{figure}

We also examine the dependence of the coupling strength on the diameter of the SWCNT.
We assume in each case that the cavity mode is tuned to be in resonance with the bright exciton energy and that the separation between the tubes is equal to $3$~nm.
In Fig.~\ref{g_d} the dependence of the coupling constant $g$ is plotted versus the diameter of the tube, the data are taken from Table~\ref{tbl}.
The dependence $g(d)$ demonstrates a universal $d^{-1/2}$ scaling. The latter follows directly from Eq.~(\ref{Rabi-2}), if one rewrites all the diameter-dependent quantities in their explicit form (see Supporting Information part B).


Finally we note that the ratio between the coupling strength and the exciton creation energy, $E_X$, is strongly enhanced compared to the case of inorganic cavities, and may reach values of $g/E_X \sim 0.1$.
This means that the off-resonant light matter coupling terms can also play a non-negligible role in the geometry considered here, and the regime of so-called ultra-strong coupling can be achieved~\cite{Ciuti2005,Liberato2014}.
We also note that in principle even higher values of $g$ can be reached if one reduces the distance between the nanotubes below $l_{min}$.
In this case however the excitonic spectrum of the system will  be substantially modified by hopping between neighbouring tubes.
Including these effects goes beyond the scope of the present paper and is left for a future investigation.


\section*{Mode dynamics in dark and bright regimes}

\subsection*{Theory}
Whilst the idea of using strong coupling to lower the energy of the bright state so that it becomes the lowest energy state of the system is attractive, it is important to assess the effectiveness of this approach. 
In the following section we develop a quantitative theoretical model to describe the mode occupancies both in the presence of the cavity and in the absence of the cavity.

First, we look at the no cavity case. In this case the coherent part of the Hamiltonian reads,
\begin{equation}
\hat{H}_{\mathrm{exc}} = E_X \hat{a}^\dagger \hat{a} +E_X^{\mathrm{dark}} \hat{b}^\dagger \hat{b}, 
\end{equation}
where $\hat{a}$, $\hat{b}$ stand for annihilation operators of bright and dark excitons, respectively. The coherent dynamics and the decay processes can be obtained from a Lindblad-type master equation for the density matrix,
\begin{equation}
i\hbar\frac{\mathrm{d}\rho}{\mathrm{d}t} =[\hat{H}_{\mathrm{exc}},\rho] + \mathcal{L}^{\mathrm{(dis)}}\rho +\mathcal{L}^{\mathrm{(th)}}\rho,
\end{equation}
where the last term represents coupling with phonons, and  $\hat{\mathcal{L}}^{\rm{(dis)}}$ is a Lindblad super-operator of the form $\hat{\mathcal{L}}^{\rm{(dis)}}\rho =\sum_i (1/\tau_i) (\hat{a}_i\rho \hat{a}_i^\dag-\{\hat{a}_i^\dag \hat{a}_i,\rho \}/2)$ with $\hat{a}_i= \hat{a}, \hat{b}$ and $\tau_i$ are the lifetimes of the modes, defined by all non-radiative decay processes. The dynamics of an arbitrary operator $\hat{O}$ in the mean-field regime reads 
$\partial \langle \hat{O}\rangle / \partial t = \langle \hat{O} \partial \rho/\partial t \rangle$, where $\langle \rangle$ denotes averaging over the density matrix. Finally, the phonon-assisted transitions between dark and bright modes within the Born-Markov approximation can be accounted for using,
\begin{equation}
\frac{\mathrm{d} \langle \hat{O} \rangle}{\mathrm{d} t} =\delta_{\delta,\hbar\omega} \frac{1}{\hbar^2\gamma_{\mathrm{ph}}} \left(  \langle [\hat{H}^+,[\hat{O},\hat{H}^-]]\rangle +\langle [\hat{H}^-,[\hat{O},\hat{H}^+]]\rangle \right),
\label{Markov}
\end{equation}
where,
\begin{equation}
\hat{H}^+ = D \sum_{\omega} \hat{a}^\dagger \hat{d}_{\omega} \hat{b}, \qquad \hat{H}^- = D \sum_{\omega} \hat{b}^\dagger \hat{d}^\dagger_{\omega} \hat{a}.
\end{equation}
Here $D$ is the exciton-phonon interaction rate, $\gamma_{\mathrm{ph}}$ describes the characteristic broadening and $\hat{d}_{\omega}$ is a phonon mode with energy $\hbar\omega$.
Note that in Eq.~(\ref{Markov}) we have applied the rotating wave approximation, retaining only scattering processes on phonons resonant with energy splitting between bright and darks exciton states ($\hbar\omega=\delta$). 

We characterize the dynamics of the system by means of coupled equations for the mode occupancies of the form,
\begin{align}
\label{exc-diff-eq}
\frac{\mathrm{d} n^0_B}{\mathrm{d} t} =P &-\frac{n^0_B}{\tau_B} +W\left(e^{-\frac{\delta}{k_B T}}n^0_D(n^0_B+1)-n^0_B(n^0_D+1)\right), \notag \\
\frac{\mathrm{d} n^0_D}{\mathrm{d} t} = &-\frac{n^0_D}{\tau_D} +W\left( n^0_B(n^0_D+1) -e^{-\frac{\delta}{k_B T}} n^0_D(n^0_B+1) \right), 
\end{align}
where $n^0_B=\langle \hat{a}^\dagger\hat{a} \rangle$, $n^0_D=\langle \hat{b}^\dagger\hat{b} \rangle$, and $W=2D^2(n_{\mathrm{ph}}+1)/(\hbar^2 \gamma_{\mathrm{ph}} )$ is the effective rate of phonon-exciton coupling.
$n_{\mathrm{ph}}=\langle \hat{d}_{\delta/\hbar}^\dagger \hat{d}_{\delta/\hbar} \rangle=1/(e^{\delta/(k_B T}-1)$ is the phonon number defined by temperature $T$. $k_B$ denotes Boltzmann constant, and $P$ corresponds to pumping the bright exciton mode, i.e. populating the bright exciton state at some rate $P$. 

The efficiency of luminescence can be characterized via the ratio of bright and dark exciton occupations, which we write as,
\begin{equation}
\label{lambda0}
\lambda^0 =\frac{n^0_B}{n^0_B+n^0_D}.
\end{equation}
The limit $\lambda^0 \rightarrow 1$ corresponds to an ideal radiative state. 

\begin{figure}
  \centering
  \includegraphics[width=0.6\linewidth]{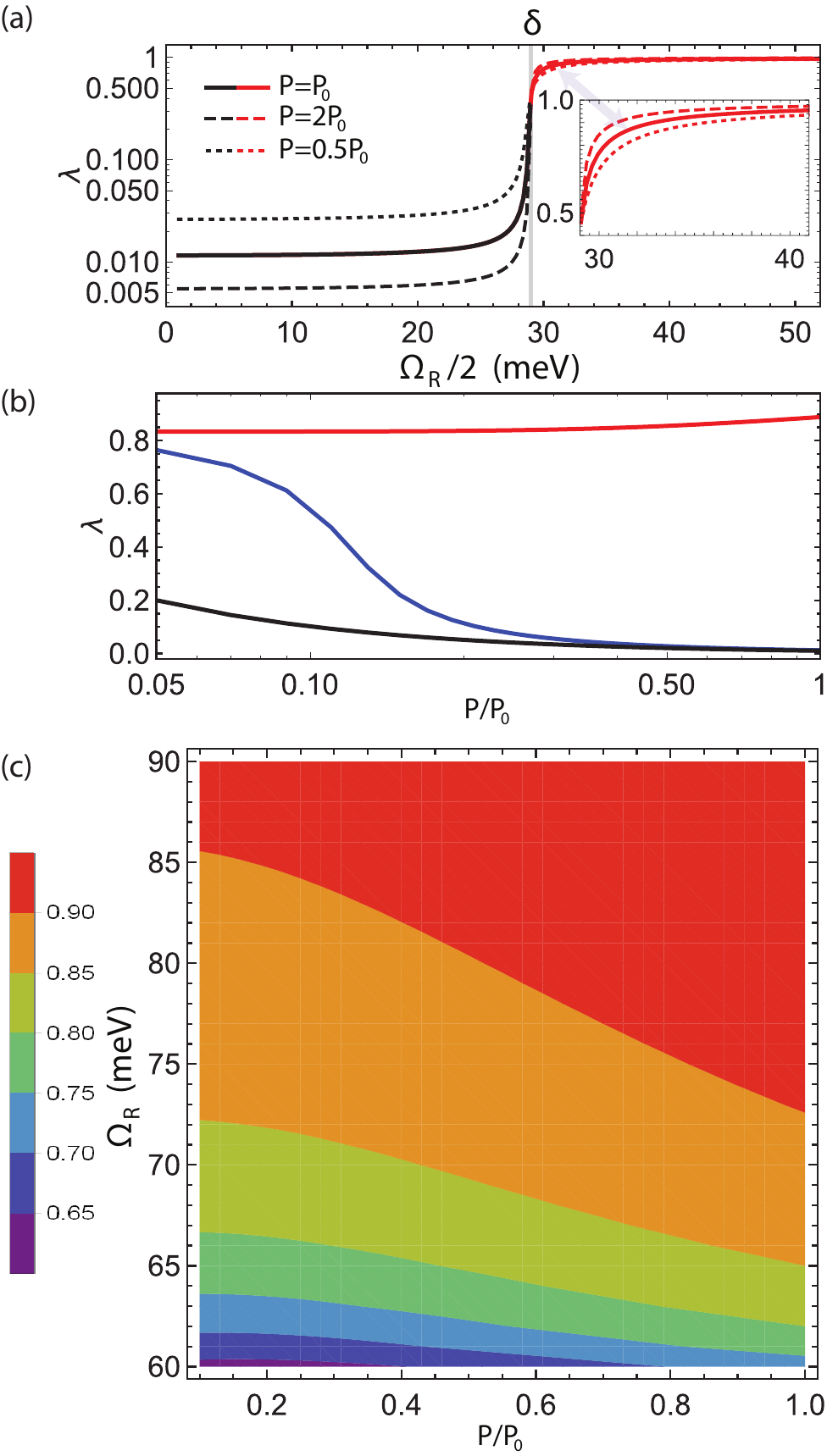}
  \caption{(a) Dependence of brightening coefficient, $\lambda$, on the Rabi splitting for various pumping levels. For smaller values of splitting, where the dark exciton is still the ground state, the effect of the cavity on the radiation efficiency is minimal. In the vicinity of the brightening area (vertical grey line, just below 30 meV) a rapid increase of radiation efficiency takes place. Here  $P_0=10^{13}$ s$^{-1}$. This population pump rate could for example be achieved optically with a resonant pump of intensity $I\approx 10$ mW/cm$^2$. (b) The pump rate dependence in the exciton (black line, no cavity), dark polariton, $\Omega_R=35$ meV (blue line) and bright polariton, $\Omega_R=70$ meV (red line) regimes. In the bright regime an increase of pump rate enhances the stimulated transition to the lower polariton branch, increasing the emission rate. In the dark regimes due to stimulated transition to dark exciton state the situation is opposite, with low pump regime being more efficient.
  (c) The brightening coefficient $\lambda$ dependence on the pump rate and Rabi splitting in a bright regime. Above threshold values of splitting and pump rate the system enters a perfectly radiative state.}
  \label{Rabi_pump_dep}
\end{figure}

Second, now that we have looked at the dynamics of the no-cavity case we proceed to discuss the polaritonic regime. In the presence of the cavity the Hamiltonian takes the form,
\begin{equation}
\hat{H}_{\mathrm{pol}} = E_X \hat{a}^\dagger \hat{a} +E_C \hat{c}^\dagger \hat{c} +\frac{g}{2} (\hat{a}^\dagger \hat{c} +\hat{c}^\dagger \hat{a})  +E_X^{\mathrm{dark}} \hat{b}^\dagger \hat{b}, 
\end{equation}
where $\hat{c}$ denotes the annihilation operator of photonic mode. Introducing the operators $\hat{a}_L=C\hat{a}-X\hat{c}$,
$\hat{a}_U=X\hat{a}+C\hat{c}$ (with $X,C= \sqrt{(1\pm\Delta/\Omega_R)/2}$ being Hopfield coefficients) one can perform a Bogoliubov transformation to the polariton basis,
\begin{equation}
\hat{H}_{\mathrm{pol}} = E_{LP} \hat{a}_L^\dagger \hat{a}_L +E_{UP} \hat{a}_U^\dagger \hat{a}_U +E_X^{\mathrm{dark}} \hat{b}^\dagger \hat{b}.
\end{equation}
Assuming the "bright" regime, i. e. $\Omega_R/2>\delta$, is achieved then the phonon-assisted transitions between the polariton branches and the dark exciton states can be characterized in the following way,
\begin{align}
\hat{H}_L^+ = C D \sum_{\omega} \hat{a}_L^\dagger \hat{d}_{\omega}^\dagger \hat{b}, \qquad \hat{H}^-_L = C D \sum_{\omega} \hat{b}^\dagger \hat{d}_{\omega} \hat{a}_L, \notag \\
\hat{H}_U^+ = X D \sum_{\omega} \hat{a}_U^\dagger \hat{d}_{\omega} \hat{b}, \qquad \hat{H}^-_U = X D \sum_{\omega} \hat{b}^\dagger \hat{d}_{\omega}^\dagger \hat{a}_U. 
\end{align}
It should be noted that we have neglected the phonon-assisted relaxation from the upper to the lower polariton branch, i.e. we have assumed that the radiative decay channels are faster. In fact phonon-assisted relaxation will ultimately redistribute the particles from the bottom of the upper branch to the bottom of the lower branch, via the reservoir. While the presence of such relaxation processes can quantitatively modify the luminescence efficiency, no qualitative changes are expected. The dynamics of the system are thus given by,
\begin{align}
\label{pol-diff-eq}
\frac{\mathrm{d} n_U}{\mathrm{d} t} =P_U -\frac{n_U}{\tau_U} &+W_{U\uparrow} n_D(n_U+1) -W_{U\downarrow} n_U(n_D+1), \notag \\
\frac{\mathrm{d} n_L}{\mathrm{d} t} =P_L -\frac{n_L}{\tau_L} &+W_{L\downarrow} n_D(n_L+1) -W_{L\uparrow} n_L(n_D+1), \notag \\
\frac{\mathrm{d} n_D}{\mathrm{d} t} = -\frac{n_D}{\tau_D} &+W_{L\uparrow} n_L(n_D+1) -W_{L\downarrow} n_D(n_L+1) \notag \\
&+W_{U\downarrow} n_U(n_D+1) -W_{D\uparrow} n_D(n_U+1), 
\end{align}
where $P_U=X^2 P$, $P_L=C^2 P$, $\tau_U^{-1}=X^2 \tau_B^{-1} + C^2 \tau_C^{-1}$, $\tau_L^{-1}=C^2 \tau_B^{-1} + X^2 \tau_C^{-1}$. 
Here the transition rate $W_{L\downarrow}=2C^2D^2(n_{ph}^L+1)/(\hbar^2 \gamma_{\mathrm{ph}} )$, where $n_{ph}^L=1/(e^{(\Omega_R/2-\delta)/(k_B T}-1)$. ($P_U$ and  $P_L$ are the population pumping rates of the upper and lower polaritons respectively.) Thus, we can express the rate $W_{L\downarrow}$ via the bare rate $W$ as,
\begin{equation}
W_{L\downarrow}=C^2 \frac{1-e^{-\delta/(k_B T)}}{1-e^{-(\Omega_R/2-\delta)/(k_B T)}} W.
\end{equation}
The remaining terms in similar fashion found as
\begin{align}
W_{U\downarrow}=X^2 \frac{1-e^{-\delta/(k_B T)}}{1-e^{-(\Omega_R/2+\delta)/(k_B T)}} W, \notag \\
W_{L\uparrow}=e^{-(\Omega_R/2-\delta)/(k_B T)} W_{L\downarrow}, \notag \\
W_{U\uparrow}=e^{-(\Omega_R/2+\delta)/(k_B T)} W_{U\downarrow}. 
\end{align}
In analogy with Eq.~(\ref{lambda0}), we introduce the characteristic luminescence efficiency in the form,
\begin{equation}
\label{lambda}
\lambda =\frac{C^2n_L+X^2n_U}{C^2n_L+X^2n_U+n_D}.
\end{equation}

Finally we note that in the case of $\Omega_R/2<\delta$, where the ground state is represented by a dark exciton (see. Fig.~\ref{sketch} (C)) for the correct description of the system the dynamical equations~(\ref{pol-diff-eq}) and corresponding transition rates should be modified (not shown).

\subsection*{Numerical simulations}


To gain further insight into the system's properties, we undertook a number of numerical simulations.
The lifetimes of the DE and BE modes can be estimated as $\tau_{D,B}=100$ ps \cite{Perebeinos2008}, and for the cavity mode $\tau_C=10$ ps.
Throughout this section we set the detuning to be $\Delta=0$ meV, and assumed room temperature operation ($T=300$K), consistent with experiments \cite{Zakharko2016}.
We analyze the dependence of the brightening coefficient on both the Rabi splitting and on the population pump rate, $P$, as found from steady-state solutions of equations (\ref{exc-diff-eq}) and (\ref{pol-diff-eq}). 

We choose $W=1.5$ ns$^{-1}$, corresponding to a brightening value of $\lambda^0=0.01$. This choice for the value of $W$ is not critical, changing it does not lead to any qualitative changes in the outcome for a wide range of $W$ values (see Supporting Information part C for details).
The pump rate is chosen as $P_0=10^{13}$ s$^{-1}$.
This population pump rate could for example be achieved optically with a resonant pump of intensity $I\approx 10$ mW/cm$^2$. 
The results of these calculations are shown in Fig. \ref{Rabi_pump_dep}. Panel (a) shows Rabi splitting dependence, accounting the region  $\Omega_R/2 < \delta$,  where the ground state is represented by a dark exciton (black curves).
The simulations show that in this regime the presence of cavity has little impact on the emission efficiency, leaving it almost at the same level.
Instead, the emission is quite sensitive to the pump rate, as discussed below.  
A sharp transition to radiative state appears when approaching the limit $\Omega_R/2\rightarrow \delta$. After entering the bright regime (red curves) we quickly reach the limit $\lambda\approx 1$.  

Another important difference between bright and dark regimes appears in the response to an increase in pump rate, plotted in the Fig. \ref{Rabi_pump_dep} (b). In both the absence of cavity and in the polaritonic dark regime (when $\Omega_R/2<\delta$), corresponding to Figs. \ref{sketch} (c), (d) respectively, all the excitation generated by pumping eventually scatters to reach the dark exciton state. Thus, increasing the pump rate reduces the radiation efficiency, as seen in Fig. \ref{Rabi_pump_dep} (b).  The opposite situation occurs in the bright regime, corresponding to Fig. \ref{sketch} (e). Here, because of the energetic configuration of levels involved, particles stay in the optically active upper and lower polariton branches, so that increasing the pump rate enhances the luminescence efficiency.

Finally, the phase diagram in panel (c) shows the brightening efficiency as a function of the pump power and the Rabi splitting. In the region of high Rabi splitting and pump rate the radiation efficiency approaches 1. Interestingly in the case of large values of Rabi splitting even at low pump rates we are still in a strongly radiative regime, which underlines the advantages of our proposed  protocol for carbon nanotube brightening, and demonstrates the potential for creation of a low-threshold radiative source. Taken together the results shown in Fig. \ref{Rabi_pump_dep} suggest that the proposed brightening scheme will be both effective and efficient.

\section*{Conclusions}
We have theoretically analyzed the regime of strong light-matter coupling in a microcavity containing an ensemble of semiconducting carbon nanotubes.
We have demonstrated that for realistic parameters of the system the value of the Rabi splitting can be greater than the splitting between bright and dark energy states.
As a result, the lower polariton mode can have an energy lower than that of the dark exciton.
It is this re-ordering of the energy of the bright and dark states which leads to a dramatic improvement of the emissive properties of nanotube-based systems.

The results of our dynamical calculations show that a significant increase in emission efficiency may be expected.
Our results are further supported by a number of very recent reports.
First, in work using strong coupling to modify inter-system crossing in the light-emitting material Erythrosine B ~\cite{Stranius2018} Stranius et al. lowered the energy of singlet level (bright) towards that of the triplet level (dark), and the authors hope to be able to push the singlet level below that of the triplet in the near future.
Second, Graf et al~\cite{Graf2017}, using SWCNTs in a cavity, achieved sufficient Rabi splitting, and did observe a 5-fold increase in emission efficiency. However, their reference (no-cavity) sample had an emission efficiency of only $\sim 0.1$ \% so that the full extent of any improvement to efficiency can not properly be gauged. Thirdly, in another very recent report, Gao et al., reported a vacuum Rabi splitting of $\sim330~\rm{meV}$ for alighted SWCNTs in a cavity\cite{Gao2018}. This splitting easily exceeds the value that we predict will be needed to achieve brightening but they did not look at emission in their report.

Finally, we note that graphene nanoribbons also have a dark excitonic ground state~\cite{Yang2007}.
Our proposed scheme for brightening should also work for nanoribbons due to the equivalence of the optical properties of both these quasi-one-dimensional structures~\cite{Portnoi2015,Saroka2017}. We further note that a similar approach might also be used to manipulate excitonic states in transition metal dichalcogenides ~\cite{Deilmann_PRB_2017_96_201113}.\\



\section*{Author Information}
{\bf Corresponding Author}\\
*E-mails: 
M.E.Portnoi@exeter.ac.uk\\
{\bf ORCID}\\
V. Shahnazaryan: 0000-0001-7892-0550 \\
V.A. Saroka: 0000-0002-8980-6611\\
I.A. Shelykh: 0000-0001-5393-821X\\
W.L. Barnes:  0000-0002-9474-5534\\
M.E. Portnoi: 0000-0001-5618-0993\\
{\bf Notes}\\
The authors declare no competing financial interest.\\
V.A. Saroka present affiliation: Center for Quantum Spintronics, Department of Physics, Norwegian University of Science and Technology, NO-7491 Trondheim, Norway

\section*{Acknowledgements}
The authors thank I. Chestnov, O. Kyriienko and C. A. Downing for valuable discussions. This work was financially supported by the EU FP7 ITN NOTEDEV (Grant No. FP7-607521), the EU H2020 RISE project CoExAN (Grant No. H2020-644076), WLB's EU ERC project (ERC-2016-AdG-742222), and by the Government of the Russian Federation through the ITMO Fellowship and Professorship Program. VS and IAS acknowledge support from mega-grant No. 14.Y26.31.0015 and goszadanie no 3.2614.2017/4.6 of the Ministry of Education and Science of the Russian Federation.  VAS and MEP acknowledge support from FP7 IRSES projects CANTOR (Grant No. FP7-612285), and InterNoM (Grant No. FP7-612624). VS and VAS thank the University of Iceland for hospitality during the work on this project.

\begin{suppinfo}
(A) Calculation of the dipole matrix element; (B) The dependence of light-matter coupling rate on CNT diameter; (C) The brightening dependence on exciton-phonon coupling rate. This material is available free of charge via the Internet at http://pubs.acs.org.
\end{suppinfo}


\bibliography{library}

\end{document}


\section*{A: Calculation of the dipole matrix element}
\label{appDipole}
In this section we present a derivation of the matrix elements of dipole transitions within the zone-folding tight-binding approximation. For graphene-based nanostructures it is convenient to deal with the matrix elements of the velocity operator, instead of the position matrix element, which allows one to express relevant physical quantities in terms of the graphene Fermi velocity $v_F$.
These two approaches, however, are absolutely equivalent in the dipole approximation~\cite{BookLandauVolIV1997}.
The two matrix elements are related by
\begin{equation}
\label{eq:VMEvsPME}
  \vec{v}_{nm} =  \dfrac{i}{\hbar} \langle n \left| \left[ H, \vec{r} \right] \right| m \rangle = i \omega_{nm} \vec{r}_{nm}\, ,
\end{equation}
where $\omega_{n m} = (E_n - E_m)/\hbar $ is the frequency of the transition.

To calculate velocity operator matrix elements for SWCNTs in the tight-binding model we employ the so-called gradient approximation (the term ``effective mass approximation'' is used in earlier literature)~\cite{Wannier1962,Blount1962a,Dresselhaus1967,LewYanVoon1993,Lin1994,Saroka2017}. The optical selection rules resulting from such a calculation are in agreement with those in Refs.~\cite{Ajiki1994,Milosevic2003,Jiang2004,Malic2006}. Using this approach it is easy to show that the velocity matrix element dependence on the wavevector is not significant in semiconducting SWCNTs in the vicinity of the Dirac point, which corresponds to the conduction and valence band edges. Therefore, the matrix element can be reasonably approximated by a constant value equal to the Fermi velocity of electrons in graphene $v_F$.
We are interested in the transitions from the edges of the closest conduction and valence subbands, i.e., those that occur near the Dirac point, we can estimate the magnitude of the dipole moment operator matrix element as follows:

\begin{equation}
\label{eq:DipoleMomentOperatorMatrixElementInTheDiracPoint}
    d_{cv} = e|r_{cv}| = \dfrac{e v_F \hbar}{E_g} = \dfrac{e \dfrac{\sqrt{3} at }{2 \hbar} \hbar}{\dfrac{2 a t}{\sqrt{3} d}} = \dfrac{3 e d}{4} \, ,
\end{equation}
where we have used Eq.~\eqref{eq:VMEvsPME} and the fact that the bandgap energy, $E_g$, in the low-energy zone-folding approximation is given by,
\begin{eqnarray}
\label{eq:CNTBandGapsInConicalApproximation}
    E_g = 2 \hbar v_F \Delta = 4 \hbar v_F /(3 d) = 2 a t /\sqrt{3} d,
\\    \nonumber
\end{eqnarray}

\noindent where $v_F = \sqrt{3} a t /(2 \hbar)$ is the graphene Fermi velocity~\cite{Neto2009} in terms of the hopping integral $t=3.033$~eV~\cite{SaitoBook1998} and $\Delta=2/(3d)$ is the shift of the momentum quantization line from the Dirac point for any semiconducting SWCNT~\cite{Saito2000,Samsonidze2003}. 

Although the majority of the results presented in this paper were obtained for a zigzag structure with chirality $(10,0)$, the approach developed here is of a general character and can be applied to semiconducting nanotubes of an arbitrary configuration.

\section*{B: The dependence of light-matter coupling rate on CNT diameter}
\label{g-d-Dep}
Here we briefly examine the CNT diameter dependence of parameters involved in the light-matter coupling rate, given by Eq.~(10) of the main text. First of all we note, that in near-resonant regime $\Delta\ll E_C(0)$, $E_C(0) \approx E_X =E_g -E_b$. As it was shown in the previous section, the bandgap scales as $E_g \propto 1/d$. From the Eq.~(6) of the main text and the following discussion one founds for the binding energy and Bohr radius: $E_b\propto 1/d$, and $a_0 \propto d$. Thus, we found that $E_C(0)\propto 1/d$. On the other hand, the distance between mirrors $L_C$ is defined through the resonant wavelength: $L_C=2\lambda_C\propto1/E_C(0)\propto d$. Hence, the factor under square root has overall $1/d^2$ dependence, which cancels out with interband transition matrix element, defined by Eq.~(\ref{eq:DipoleMomentOperatorMatrixElementInTheDiracPoint}). Thus, what it remains is to examine the exciton wave function at zero interparticle distance, reading as $\psi_{\alpha}(0) = N_{\alpha} W_{\alpha,1/2}(2\gamma d/\alpha a_0)$, where
\begin{equation}
    N_{\alpha}=\left( \alpha a_0 \int\limits_{2\gamma d/\alpha a_0}^\infty |W_{\alpha,1/2}(z)|^2 \mathrm{d} z \right) ^{-1/2}. 
\end{equation}
As it follows from the definition of Bohr radius, the factor $2\gamma d/\alpha a_0$ do not depend on tube diameter $d$. Finally, the last parameter depending on $d$ is the normalization constant $N_\alpha \propto d^{-1/2}$ via the Bohr radius $a_0$. 

\section*{C: The brightening dependence on exciton-phonon coupling rate}
\label{appRate}

\begin{figure}[h]
	\includegraphics[width=0.9\linewidth]{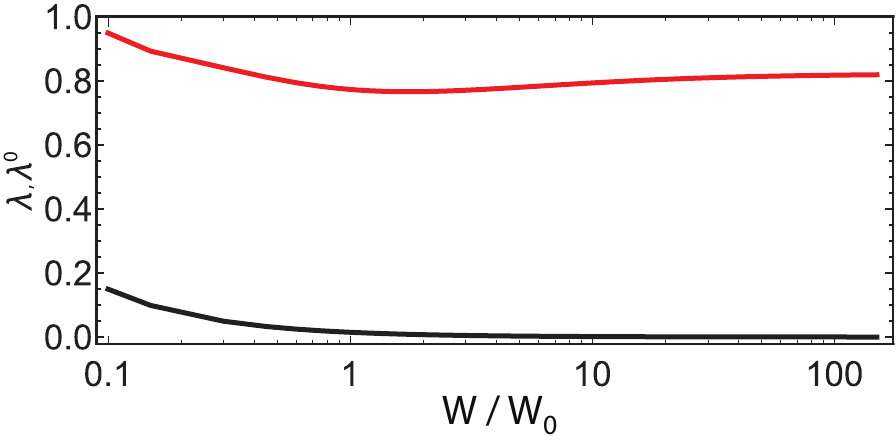}
    \caption{The luminescence rates $\lambda$, $\lambda^0$ versus exciton-phonon coupling rate $W$. The black line corresponds to the dark regime ($\lambda^0$) and the red line to the bright regime ($\lambda$). Here $W_0 =1.5$ ns $^{-1}$ is the value used in the main text. The other parameters are $P=P_0$ and $\Omega_R=60$ meV.}
\label{fig:W_dep}
\end{figure}

In the  following section we discuss the impact of the exciton-phonon effective coupling rate $W$ on the efficiency of the luminescence in both bright and dark regimes. Particularly, in Fig. \ref{fig:W_dep} we plot the luminescence efficiencies $\lambda^0$, $\lambda$ as a function of $W$. We choose a logarithmic scale for the $x$ axis, varying $W$ over several orders of magnitude. 
While the value of luminescence rate in the excitonic regime (i.e. in the absence of a cavity) varies with the coupling rate $W$, it always remains in the low-efficiency regime (black line).
On the other hand, over the whole range of $W$ plotted, the impact of the cavity strikingly increases the luminescence rate (red line).

\bibliography{library}